\documentclass[11pt]{article}
\usepackage{epsfig} 
\newcommand{\Dslash}{D \! \! \! \! /}

\newcommand{\half}{\mbox{\small{$\frac{1}{2}$}}} 
 
\newcommand{\Nf}{N_{\!f}} 
\newcommand{\MSbar}{\overline{\mbox{MS}}} 

\begin{document}
\title{Three loop $\MSbar$ renormalization of the Curci-Ferrari model and the 
dimension two BRST invariant composite operator in QCD} 
\author{J.A. Gracey, \\ Theoretical Physics Division, \\ 
Department of Mathematical Sciences, \\ University of Liverpool, \\ Peach 
Street, \\ Liverpool, \\ L69 7ZF, \\ United Kingdom.} 
\date{} 
\maketitle 
\vspace{5cm} 
\noindent 
{\bf Abstract.} The massless Curci-Ferrari model with $\Nf$ flavours of quarks 
is renormalized to three loops in the $\MSbar$ scheme in an arbitrary covariant
gauge with parameter $\alpha$. The renormalization of the BRST invariant 
dimension two composite operator, $\half A^{a\,2}_\mu$ $-$ $\alpha \bar{c}^a
c^a$, which corresponds to the mass operator in the massive Curci-Ferrari 
model, is determined by renormalizing the Green's function where the operator 
is inserted in a ghost two-point function. Consequently the anomalous dimension
of the QCD Landau gauge operator, $\half A^{a\,2}_\mu$, and the (gauge 
independent) photon mass anomalous dimension in QED are both deduced at three 
loops. 

\vspace{-17.5cm}
\hspace{13.5cm}
{\bf LTH 565}

\newpage 
There has been a renewed interest recently in trying to understand the origin 
of confinement in Yang-Mills theories and QCD from the point of view of the
existence of a non-zero vacuum expectation value of the dimension two composite
operator $\half A^{a \, 2}_\mu$ where $A^a_\mu$ is the gluon field. For 
instance, lattice gauge theory studies by Boucaud et al, for example, \cite{1}
and references therein, have shown evidence that this operator has a non-zero 
vacuum expectation value. Whilst it might appear that the operator itself can 
have no physical relevance due to its lack of gauge invariance, it has been 
argued for instance in, \cite{2}, that in the Landau gauge it is gauge 
invariant since it is one term of a more general dimension two operator which 
is non-local in non-Landau gauges. With the assumption that the non-zero vacuum 
expectation value $\langle \half A^{a \, 2}_\mu \rangle$ is present in 
Yang-Mills theories it has been the subject of various analytic and field 
theoretic investigations either using the operator product expansion, 
\cite{3,4,5}, or other methods, \cite{2,6,7,8,9,10,11,12,13}. One interesting
approach has been that of \cite{2}. There the effective potential of the 
composite operator itself is computed at two loops in $SU(N)$ Yang-Mills 
theory. It is demonstrated that the associated auxiliary field develops a 
non-zero vacuum expectation value in the true vacuum. The classical vacuum
where the vacuum expectation value remains zero is energetically unstable. This
calculation developed the early work of \cite{14,15} to compute the effective
potential of a composite operator in a field theory to two loops which is a 
non-trivial exercise. One of the reasons for this is that non-perturbative 
terms arise at leading order due to the presence of a $1/g^2$ term where $g$ is
the basic coupling constant. The upshot of this is that one has to compute 
several quantities to {\em three} loops including the anomalous dimension of 
the gluonic dimension two local composite operator, \cite{2}. This was achieved
by the tensor correction technique developed in \cite{16} to handle massive and 
massless Feynman integrals to three loop order in automatic Feynman diagram 
computer programmes. For example, the method of \cite{2} produced a numerical 
estimate for the vacuum expectation value of $\half A^{a \, 2}_\mu$ in 
Yang-Mills theory in the context of this essentially perturbative probing of a 
non-perturbative phenomenon. Clearly those non-perturbative contributions from 
instantons are not included in this approach but this does not detract from its
success and potential application to other situations. Indeed in this context 
there has been recent studies of similar vacuum expectation value generation 
problems in Yang-Mills theories in different gauges. For instance, in 
\cite{6,7,8,9,10,11,12,13} the condensation of ghost number breaking vacuum 
expectation values has been investigated at one loop in the Landau gauge as 
well as the more interesting maximal abelian gauge. In the latter case the 
off-diagonal ghosts gain a mass in contrast to the diagonal ghosts remaining 
massless. One hope is that the same feature occurs in the gluon sector, 
indicating that the centre of the group is special for confinement since 
abelian monopoles are believed to drive this mechanism.  

Given this current interest in this area it is worth noting that it can be
pursued in several directions. Clearly a two loop extension of \cite{12,13}, 
for instance, would be interesting. However, all the current investigations 
have been for Yang-Mills theories. For more realistic studies of the effective 
potential approach one needs to include $\Nf$ flavours of quarks. Therefore, 
the aim of this article is to provide the first stage in this problem which is 
the determination of the three loop anomalous dimension of the 
$\half A^{a \, 2}_\mu$ composite operator including quarks, thereby extending
the result given in \cite{2}. To achieve this we will in fact deduce it as a 
consequence of the renormalization of a model similar to QCD in its ultraviolet
properties but differing from it in the infrared. Known as the Curci-Ferrari 
model, \cite{17}, it was believed it could be central to understanding massive 
vector bosons as an alternative to the Higgs mechanism. However, as it is also 
not a unitary model, \cite{18,19,20,21,22,23} it has only received renewed 
interest due to its relation to the ghost condensation problem since the 
Curci-Ferrari model has a feature similar to the maximal abelian gauge and
QCD in a class of nonlinear gauges, \cite{24,25}, which is the presence of a
four ghost interaction which is a crucial ingredient for the phenomenon. As the
Curci-Ferrari model has a natural dimension two BRST invariant mass term it can
be used to determine the ultraviolet structure of the renormalization group 
functions of QCD itself since the mass acts as an infrared regulator, 
\cite{17,20,21,22,23,26,27}. The model is renormalizable and the 
renormalization group functions are in agreement with those of QCD in the 
Landau gauge. As this BRST dimension two operator is the one required for the 
earlier discussion we will determine its anomalous dimension in the 
Curci-Ferrari model in an arbitrary covariant gauge and then specialize to the 
Landau gauge. Whilst this may appear a roundabout method we can exploit several
technical points to ease our computation. Moreover, we will additionally 
renormalize the Curci-Ferrari model itself at three loops and extend the lower 
order calculations of \cite{17,21,28,29} thereby addressing several problems
simultaneously. 

We begin by recalling the Curci-Ferrari model which includes a BRST invariant
mass term for the gluon. The Lagrangian is, \cite{17},  
\begin{eqnarray} 
L^{\mbox{\footnotesize{CF}}} &=& -~ \frac{1}{4} G_{\mu\nu}^a 
G^{a \, \mu\nu} ~-~ \frac{1}{2\alpha} (\partial^\mu A^a_\mu)^2 ~+~ 
\frac{1}{2} m^2 A_\mu^a A^{a \, \mu} ~-~ \bar{c}^a \partial^\mu D_\mu c^a ~-~ 
\alpha m^2 \bar{c}^a c^a \nonumber \\
&& +~ \frac{g}{2} f^{abc} \partial^\mu A^a_\mu \, \bar{c}^b c^c ~+~ 
\frac{\alpha g^2}{8} f^{eab} f^{ecd} \bar{c}^a c^b \bar{c}^c c^d ~+~ 
i \bar{\psi}^{iI} \Dslash \psi^{iI} ~-~ \sqrt{\beta} m \bar{\psi}^{iI} 
\psi^{iI} 
\label{cflag} 
\end{eqnarray}  
where $A^a_\mu$ is the gluon field, $c^a$ and $\bar{c}^a$ are the ghost and 
antighost fields, $\psi^{iI}$ is the quark field and $\alpha$ is the covariant 
gauge fixing parameter. The covariant derivatives are given by 
$D_\mu \psi$~$=$~$\partial_\mu \psi$~$+$~$ig A^a_\mu T^a \psi$ and 
$D_\mu c^a$~$=$~$\partial_\mu c^a$~$-$~$g f^{abc} A^b_\mu c^c$ implying 
$G^a_{\mu\nu}$ $=$ $\partial_\mu A^a_\nu$ $-$ $\partial_\nu A^a_\mu$ $-$ 
$g f^{abc} A^b_\mu A^c_\nu$ where $f^{abc}$ are the structure constants of the
colour group whose generators are $T^a$, $1$~$\leq$~$a$~$\leq$~$N_A$, 
$1$ $\leq$ $I$ $\leq$ $N_F$ and $1$ $\leq$ $i$ $\leq$ $\Nf$ with $N_A$ and 
$N_F$ are the dimensions of the respective adjoint and fundamental 
representations. The gluon mass is $m$ and the quark mass is expressed in terms
of this basic scale with the parameter $\beta$ introduced to indicate that the 
masses are different. The model is renormalizable and the renormalization group
functions are known at two loops, \cite{28,29}. In the case when $m^2$ $=$ $0$ 
one has QCD fixed in a nonlinear gauge which unlike the Curci-Ferrari model is 
a unitary theory. The presence of the non-zero mass in (\ref{cflag}) breaks 
both unitarity and the nilpotency of the BRST transformation, 
\cite{18,19,20,21,22,23,24,25}. To compute the renormalization constants of 
(\ref{cflag}) we follow a two stage approach. First, we determine the basic 
renormalization constants of the fields and parameters and then deduce the 
renormalization of the mass operator treated as a composite operator inserted 
in a Green's function based on the {\em massless} Lagrangian. This method has 
been used, for example, to determine the renormalization of the quark mass in 
QCD at three loops in \cite{30,31}. The advantage of considering the massless 
Lagrangian is that one can apply the {\sc Mincer} algorithm \cite{32}, as 
written in version $2.0$ of the symbolic manipulation language {\sc Form}, 
\cite{33,34}. This is an efficient automatic Feynman diagram package which 
determines the ultraviolet structure of massless three loop two point functions
with respect to dimensional regularization in $d$ $=$ $4$ $-$ $2\epsilon$ 
dimensions. More specifically to achieve this for each of the Green's functions
we need to examine, we generate the appropriate set of Feynman diagrams using 
the {\sc Qgraf} package, \cite{35}, in a format which is readable in 
{\sc Form}, \cite{34}. These are then integrated using the appropriate 
{\sc Mincer} routine after the basic topology of the diagrams has been 
identified. As the first stage in the computation concerns the wave function 
renormalization which derive from the two point functions we note that the 
renormalization constants are defined by 
\begin{eqnarray} 
A^{a \, \mu}_{\mbox{\footnotesize{o}}} &=& \sqrt{Z_A} \, A^{a \, \mu} ~~,~~ 
c^a_{\mbox{\footnotesize{o}}} ~=~ \sqrt{Z_c} \, c^a ~~,~~ 
\bar{c}^a_{\mbox{\footnotesize{o}}} ~=~ \sqrt{Z_c} \, \bar{c}^a ~~,~~ 
\psi_{\mbox{\footnotesize{o}}} ~=~ \sqrt{Z_\psi} \psi \nonumber \\ 
g_{\mbox{\footnotesize{o}}} &=& Z_g \, g ~~,~~ m_{\mbox{\footnotesize{o}}} ~=~ 
Z_m \, m ~~,~~ \alpha_{\mbox{\footnotesize{o}}} ~=~ Z^{-1}_\alpha Z_A \, 
\alpha ~~,~~ \beta_{\mbox{\footnotesize{o}}} ~=~ Z_\beta \beta ~.  
\label{Zdef}
\end{eqnarray} 
where the subscript ${}_{\mbox{\footnotesize{o}}}$ denotes the bare quantity.
We have, using the modified minimal subtraction scheme,  
\begin{eqnarray} 
Z_A &=& 1 ~+~ \left[ \left( \frac{13}{6} - \frac{\alpha}{2} \right) C_A 
- \frac{4}{3} T_F \Nf \right] \frac{a}{\epsilon} \nonumber \\
&& +~ \left[ \left( \left( \frac{3\alpha^2}{16} - \frac{17\alpha}{24} 
- \frac{13}{8} \right) C_A^2 + C_A T_F\Nf \left( \frac{2}{3}\alpha + 1 \right) 
\right) \frac{1}{\epsilon^2} \right. \nonumber \\
&& \left. ~~~~~-~ \left( \left( \frac{\alpha^2}{16} + \frac{11\alpha}{16}
- \frac{59}{16} \right) C_A^2 + 2 C_F T_F\Nf + \frac{5}{2} C_A T_F \Nf \right) 
\frac{1}{\epsilon} \right] a^2 \nonumber \\ 
&& +~ \left[ \left( \left( \frac{403}{144} + \frac{47\alpha}{48} 
+ \frac{\alpha^2}{8} - \frac{\alpha^3}{16} \right) C_A^3 
- C_A^2 T_F\Nf \left( \frac{22}{9} + \frac{5\alpha}{6} + \frac{\alpha^2}{4} 
\right) + \frac{4}{9} C_A T_F^2 \Nf^2 \right) \frac{1}{\epsilon^3} \right. 
\nonumber \\
&& \left. ~~~~+~ \left( \left( \frac{5\alpha^3}{96} + \frac{97\alpha^2}{192} 
- \frac{143\alpha}{96} - \frac{7957}{864} \right) C_A^3 
+ C_A^2 T_F\Nf \left( \frac{481}{54} + \frac{19\alpha}{12} 
+ \frac{\alpha^2}{12} \right) \right. \right. \nonumber \\
&& \left. \left. ~~~~~~~~~~~~-~ \frac{50}{27} C_A T_F^2 \Nf^2 
- \frac{8}{9} C_F T_F^2 \Nf^2 + C_A C_F T_F\Nf \left( \frac{31}{9} + \alpha 
\right) \right) \frac{1}{\epsilon^2} \right. \nonumber \\
&& \left. ~~~~+~ \left( \left( \frac{9965}{864} - \frac{167\alpha}{96} 
- \frac{101\alpha^2}{384} - \frac{\alpha^3}{64} - \left( \frac{\alpha}{4} 
+ \frac{3}{16} \right) \zeta(3) \right) C_A^3 \right. \right. \nonumber \\ 
&& \left. \left. ~~~~-~ C_A^2 T_F\Nf \left( \frac{911}{54} - 6\zeta(3) 
- \frac{2\alpha}{3} \right) 
- C_A C_F T_F\Nf \left( \frac{5}{54} + 8\zeta(3) \right) 
+ \frac{76}{27} C_A T_F^2 \Nf^2 \right. \right. \nonumber \\
&& \left. \left. ~~~~+~ \frac{44}{27} C_F T_F^2 \Nf^2 
+ \frac{2}{3} C_F^2 T_F \Nf \right) \frac{1}{\epsilon} \right] a^3 ~+~ O(a^4) 
\label{Za} 
\end{eqnarray} 
\begin{eqnarray} 
Z_\alpha &=& 1 ~-~ \left( \frac{\alpha}{4} \right) C_A \frac{a}{\epsilon} ~+~ 
C_A^2 \left[ \left( \frac{\alpha^2}{16} + \frac{3\alpha}{16} 
\right) \frac{1}{\epsilon^2} ~-~ \left( \frac{\alpha^2}{32} 
+ \frac{5\alpha}{32} \right) \frac{1}{\epsilon} \right] a^2 
\nonumber \\ 
&& -~ \left[ \left( \left( \frac{31\alpha}{96} + \frac{3\alpha^2}{32} 
+ \frac{\alpha^3}{64} \right) C_A^3 - \frac{\alpha}{12} C_A^2 T_F\Nf \right) 
\frac{1}{\epsilon^3} \right. \nonumber \\
&& \left. ~~~~-~ \left( \left( \frac{7\alpha^3}{384} + \frac{11\alpha^2}{64} 
+ \frac{115\alpha}{192} \right) C_A^3 - \frac{5\alpha}{24} C_A^2 T_F\Nf 
\right) \frac{1}{\epsilon^2} \right. \nonumber \\
&& \left. ~~~~+~ \left( \left( \frac{67\alpha}{128} + \frac{13\alpha^2}{128}
+ \frac{\alpha^3}{128} \right) C_A^3 - \frac{5\alpha}{16} C_A^2 T_F\Nf \right) 
\frac{1}{\epsilon} \right] a^3 ~+~ O(a^4) 
\end{eqnarray} 
\begin{eqnarray} 
Z_c &=& 1 ~+~ \left( \frac{3}{4} - \frac{\alpha}{4} \right) C_A 
\frac{a}{\epsilon} ~+~ \left[ \left( \left( \frac{\alpha^2}{16} - \frac{35}{32}
\right) C_A^2 + \frac{1}{2} C_A T_F \Nf \right) \frac{1}{\epsilon^2} \right. 
\nonumber \\ 
&& \left. ~~~~~~~~~~~~~~~-~ \left( \left( \frac{\alpha^2}{32} 
- \frac{\alpha}{32} - \frac{95}{96} \right) C_A^2 + \frac{5}{12} C_A T_F \Nf 
\right) \frac{1}{\epsilon} \right] a^2 \nonumber \\ 
&& +~ \left[ \left( \left( \frac{2765}{1152} + \frac{35\alpha}{384} 
- \frac{3\alpha^2}{64} - \frac{\alpha^3}{64} \right) C_A^3 
- C_A^2 T_F\Nf \left( \frac{149}{72} + \frac{\alpha}{24} \right) 
+ \frac{4}{9} C_A T_F^2 \Nf^2 \right) \frac{1}{\epsilon^3} \right. \nonumber \\
&& \left. ~~~~+~ \left( \left( \frac{7\alpha^3}{384} + \frac{11\alpha^2}{128} 
+ \frac{5\alpha}{96} - \frac{15587}{3456} \right) C_A^3 
+ C_A^2 T_F\Nf \left( \frac{1405}{432} - \frac{\alpha}{48} \right) 
\right. \right. \nonumber \\
&& \left. \left. ~~~~~~~~~~~~+~ C_A C_F T_F \Nf 
- \frac{10}{27} C_A T_F^2 \Nf^2 \right) \frac{1}{\epsilon^2} \right. 
\nonumber \\
&& \left. ~~~~+~ \left( \left( \frac{15817}{5184} - \frac{17\alpha}{96} 
- \frac{55\alpha^2}{768} - \frac{\alpha^3}{128} + \left( \frac{\alpha}{8} 
+ \frac{3}{32} \right) \zeta(3) \right) C_A^3 \right. \right. \nonumber \\ 
&& \left. \left. ~~~~~~~~~~~~-~ C_A^2 T_F\Nf \left( \frac{97}{324} 
+ 3\zeta(3) - \frac{7\alpha}{24} \right) - C_A C_F T_F\Nf 
\left( \frac{15}{4} - 4\zeta(3) \right) \right. \right. \nonumber \\
&& \left. \left. ~~~~~~~~~~~~-~ \frac{35}{81} C_A T_F^2 \Nf^2 \right) 
\frac{1}{\epsilon} \right] a^3 ~+~ O(a^4) 
\end{eqnarray} 
\begin{eqnarray} 
Z_\psi &=& 1 ~-~ \alpha C_F \frac{a}{\epsilon} ~+~ \left[ \left( C_F C_A  
\left( \frac{\alpha^2}{8} + \frac{3\alpha}{4} \right) + \frac{\alpha^2}{2} 
C_F^2 \right) \frac{1}{\epsilon^2} \right. \nonumber \\ 
&& \left. ~~~~~~~~~~~~~-~ \left( C_F C_A \left( \alpha + \frac{25}{8} 
\right) ~-~ C_F T_F \Nf ~-~ \frac{3}{4} C_F^2 \right) \frac{1}{\epsilon} 
\right] a^2 \nonumber \\ 
&& +~ \left[ \left( \frac{\alpha}{3} C_A C_F T_F \Nf 
- \left( \frac{3\alpha^2}{4} + \frac{\alpha^3}{8} \right) C_A C_F^2 
- C_A^2 C_F \left( \frac{31\alpha}{24} + \frac{3\alpha^2}{16} 
+ \frac{\alpha^3}{48} \right) - \frac{\alpha}{6} C_F^3 \right) 
\frac{1}{\epsilon^3} \right. \nonumber \\ 
&& \left. ~~~~+~ \left( \frac{8}{9} C_F T_F^2 \Nf^2 - \frac{3\alpha}{4} C_F^3 
+ \left( \frac{2}{3} - \alpha \right) C_F^2 T_F \Nf 
- \left( \frac{47}{9} + \alpha \right) C_A C_F T_F \Nf \right. \right. 
\nonumber \\
&& \left. \left. ~~~~~~~~~~~~+~ \left( \alpha^2 + \frac{25\alpha}{8} 
- \frac{11}{6} \right) C_A C_F^2 + \left( \frac{275}{36} + \frac{73\alpha}{24}
+ \frac{7\alpha^2}{16} + \frac{\alpha^3}{48} \right) C_A^2 C_F \right) 
\frac{1}{\epsilon^2} \right.  \nonumber \\  
&& \left. ~~~~+~ \left( - \frac{20}{27} C_F T_F^2 \Nf^2 
- \frac{1}{2} C_F^3 - C_F^2 T_F \Nf + \left( \frac{287}{27} 
+ \frac{17\alpha}{12} \right) C_A C_F T_F \Nf \right. \right. \nonumber \\
&& \left. \left. ~~~~~~~~~~~~-~ \left( \frac{\alpha^3}{32} 
+ \frac{5\alpha^2}{16} + \frac{263\alpha}{96} + \frac{9155}{432} 
- \left( \frac{23}{8} - \frac{\alpha}{4} \right) \zeta(3) \right) C_A^2 C_F 
\right. \right. \nonumber \\
&& \left. \left. ~~~~~~~~~~~~+~ \left( \frac{143}{12} - 4 \zeta(3) \right) 
C_A C_F^2 \right) \frac{1}{\epsilon} \right] a^3 ~+~ O(a^4) 
\label{Zwfs} 
\end{eqnarray} 
where $T^a T^a$ $=$ $C_F I$, $\mbox{Tr} \left( T^a T^b \right)$ $=$ $T_F
\delta^{ab}$, $f^{acd} f^{bcd}$ $=$ $C_A \delta^{ab}$, $\zeta(n)$ is the
Riemann zeta function and $a$ $=$ $g^2/(16\pi^2)$. We have checked our routines
and programmes which determine these expressions by first running the files for
the usual QCD Lagrangian as input before replacing the file defining the 
Feynman rules with those for the Curci-Ferrari model. We found exact agreement 
with all known QCD expressions for arbitrary $\alpha$, including the more 
recently determined ghost anomalous dimension, \cite{36,37,38,39,40}. In all 
our calculations of the renormalization constants we follow the technique of 
\cite{39} by computing the Green's function in terms of {\em bare} parameters 
and then rescaling them at the end via (\ref{Zdef}) which is equivalent to the
subtraction procedure. The remaining infinities are removed by fixing the 
associated renormalization constant for that particular Green's function.  

To proceed further we need to confirm that with these wave function anomalous
dimensions, the correct coupling constant renormalization emerges in the
Curci-Ferrari model. As this ought to be gauge independent for all $\alpha$
this will provide a stringent check on (\ref{Za}) to (\ref{Zwfs}). However, to 
achieve this one must consider several of the three point vertices of 
(\ref{cflag}) which, to apply the {\sc Mincer} algorithm, requires an external 
leg momentum to be nullified. In this case there is the potential difficulty 
that a spurious infrared infinity can emerge in the answer due to infrared 
infinities at the nullified vertex. Therefore, it is appropriate to either 
consider those Green's functions where this problem never arises in the first 
place or introduce the infrared rearrangement procedure, \cite{41,42}, which is
difficult to automate. We have chosen the former which can be made automatic 
for the coupling constant renormalization but note that the method will also be
central in deducing the gluon mass renormalization. We have computed the 
quark-gluon and ghost-gluon vertex renormalization at three loops and found 
that for {\em both} QCD and the Curci-Ferrari model the same gauge invariant 
coupling constant renormalization constant emerged which agrees with the 
original result of \cite{43}. We found  
\begin{eqnarray} 
Z_g &=& 1 ~+~ \left( \frac{2}{3} T_F \Nf - \frac{11}{6} C_A \right) 
\frac{a}{\epsilon} ~+~ \left[ \left( \frac{121}{24}C_A^2 + \frac{2}{3} T_F^2
\Nf^2 - \frac{11}{3} C_A T_F \Nf \right) \frac{1}{\epsilon^2} \right. 
\nonumber \\ 
&& \left. +~ \left( C_F T_F \Nf + \frac{5}{3} C_A T_F \Nf - \frac{17}{6} 
C_A^2 \right) \frac{1}{\epsilon} \right] a^2 \nonumber \\  
&& +~ \left[ \left( \frac{605}{36} C_A^2 T_F \Nf - \frac{55}{9} C_A T_F^2 \Nf^2
+ \frac{20}{27} T_F^3 \Nf^3 - \frac{6655}{432} C_A^3 \right) 
\frac{1}{\epsilon^3} \right. \nonumber \\ 
&& \left. ~~~~+~ \left( \frac{22}{9} C_F T_F^2 \Nf^2 - \frac{121}{18} 
C_A C_F T_F \Nf - \frac{979}{54} C_A^2 T_F \Nf + \frac{110}{27} C_A T_F^2 \Nf^2
+ \frac{2057}{108} C_A^3 \right) \frac{1}{\epsilon^2} \right. \nonumber \\
&& \left. ~~~~+~ \left( \frac{205}{54} C_A C_F T_F \Nf 
- \frac{22}{27} C_F T_F^2 \Nf^2 + \frac{1415}{162} C_A^2 T_F \Nf 
\right. \right. \nonumber \\ 
&& \left. \left. ~~~~~~~~~~~~-~ \frac{79}{81} C_A T_F^2 \Nf^2 
- \frac{1}{3} C_F^2 T_F \Nf - \frac{2857}{324} C_A^3 \right) \frac{1}{\epsilon}
\right] a^3 ~+~ O(a^4) ~. 
\end{eqnarray}  
This together with (\ref{Zwfs}) means the basic renormalization group functions
for the {\em massless} Curci-Ferrari model at three loops are 
\begin{eqnarray} 
\gamma_A(a) &=& \left[ ( 3\alpha - 13 ) C_A + 8T_F \Nf \right] \frac{a}{6} 
\nonumber \\
&& +~ \left[ \left( \alpha^2 + 11\alpha - 59 \right) C_A^2 + 40 C_A T_F \Nf 
+ 32 C_F T_F \Nf \right] \frac{a^2}{8} \nonumber \\  
&& +~ \left[ \left( 54\alpha^3 + 909\alpha^2 + ( 6012 + 864\zeta(3) )\alpha 
+ 648\zeta(3) - 39860 \right) C_A^3 \right. \nonumber \\ 
&& \left. ~~~~~-~ \left( 2304\alpha + 20736\zeta(3) - 58304 \right) 
C_A^2 T_F \Nf + \left( 27648\zeta(3) + 320 \right) C_A C_F T_F \Nf \right. 
\nonumber \\ 
&& \left. ~~~~~-~ 9728 C_A T_F^2 \Nf^2 - 2304 C_F^2 T_F \Nf 
- 5632 C_F T_F^2 \Nf^2 \right] \frac{a^3}{1152} ~+~ O(a^4) \nonumber \\  
\gamma_\alpha(a) &=& -~ \left[ ( 3\alpha - 26 ) C_A + 16 T_F \Nf \right]
\frac{a}{12} \nonumber \\
&& -~ \left[ \left( \alpha^2 + 17\alpha - 118 \right) C_A^2 + 80 C_A T_F \Nf 
+ 64 C_F T_F \Nf \right] \frac{a^2}{16} \nonumber \\ 
&& -~ \left[ \left( 27\alpha^3 + 558\alpha^2 + ( 4203 + 864\zeta(3) )\alpha 
+ ( 648\zeta(3) - 39860 ) \right) C_A^3 \right. \nonumber \\ 
&& \left. ~~~~~-~ \left( 1224\alpha + 20736\zeta(3) - 58304 \right) 
C_A^2 T_F \Nf + \left( 27648\zeta(3) + 320 \right) C_A C_F T_F \Nf \right. 
\nonumber \\ 
&& \left. ~~~~~-~ 9728 C_A T_F^2 \Nf^2 - 2304 C_F^2 T_F \Nf 
- 5632 C_F T_F^2 \Nf^2 \right] \frac{a^3}{1152} ~+~ O(a^4) \nonumber \\  
\gamma_c(a) &=& ( \alpha - 3 ) C_A \frac{a}{4} ~+~ \left[ \left( 3\alpha^2 
- 3\alpha - 95 \right) C_A^2 + 40 C_A T_F \Nf \right] \frac{a^2}{48} 
\nonumber \\  
&& +~ \left[ \left( 162\alpha^3 + 1485\alpha^2 + ( 3672 - 2592\zeta(3) )\alpha 
- ( 1944\zeta(3) + 63268 ) \right) C_A^3 \right. \nonumber \\ 
&& \left. ~~~~~-~ \left( 6048\alpha - 62208\zeta(3) - 6208 \right) 
C_A^2 T_F \Nf - \left( 82944\zeta(3) - 77760 \right) C_A C_F T_F \Nf \right. 
\nonumber \\ 
&& \left. ~~~~~+~ 9216 C_A T_F^2 \Nf^2 \right] \frac{a^3}{6912} ~+~ O(a^4) 
\nonumber \\  
\beta(a) &=& -~ \left[ \frac{11}{3} C_A - \frac{4}{3} T_F \Nf \right] a^2 ~-~ 
\left[ \frac{34}{3} C_A^2 - 4 C_F T_F \Nf - \frac{20}{3} C_A T_F \Nf \right]
a^3 \nonumber \\  
&& +~ \left[ 2830 C_A^2 T_F \Nf - 2857 C_A^3 + 1230 C_A C_F T_F \Nf 
- 316 C_A T_F^2 \Nf^2 \right. \nonumber \\ 
&& \left. ~~~~~-~ 108 C_F^2 T_F \Nf - 264 C_F T_F^2 \Nf^2 \right] 
\frac{a^4}{54} ~+~ O(a^5) ~.  
\label{cfrge} 
\end{eqnarray} 
It is worth noting that as a consequence we have verified the three loop gauge 
independence of the $\beta$-function by computing in this particular nonlinear 
covariant gauge. We have chosen not to re-compute the quark mass anomalous 
dimension since from the two loop calculation, \cite{29}, it is clear that the 
same three loop gauge independent expression as \cite{30,31} will emerge. For 
completeness, we note the anomalous dimension of the gauge parameter, $\alpha$,
in our conventions, (\ref{Zdef}), is  
\begin{eqnarray} 
\gamma_A(a) ~+~ \gamma_\alpha(a) &=& \alpha \left[ \frac{a}{4} C_A ~+~ 
\left( \alpha + 5 \right) C_A^2 \frac{a^2}{16} \right. \nonumber \\
&& \left. ~~~~+~ 3 C_A^2 \left[ \left( \alpha^2 + 13\alpha + 67\alpha \right) 
C_A - 40 T_F \Nf \right] \frac{a^3}{128} \right] ~+~ O(a^4) 
\end{eqnarray} 
which implies that the Landau gauge remains as a fixed point of this 
renormalization group function at three loops. 

Armed with these basic renormalization constants we have deduced the mass
renormalization by considering the composite operator 
\begin{equation} 
{\cal O} ~=~ \frac{1}{2} A^a_\mu A^{a \, \mu} ~-~ \alpha \bar{c}^a c^a
\end{equation} 
in the massless Curci-Ferrari model and renormalizing it by inserting it in an
appropriate two point function. In \cite{44} it was verified that this operator
is multiplicatively renormalizable at two loops where the one loop check was
established in \cite{5}. We expect this is an all orders property intimately 
related to the fact that it is BRST invariant. Indeed the interplay of 
renormalizability and BRST invariance of this operator has been explored at 
two loops in \cite{44}. Clearly, we need to be careful which two point function 
${\cal O}$ is inserted into, due to the problems noted earlier. There is an 
additional potential difficulty in this case in that the operator must not be 
inserted at zero momentum. In other words a momentum must flow through the 
operator, otherwise an incorrect result could be obtained for the operator
anomalous dimension. An excellent example of such pitfalls has been elegantly
expounded in the context of the {\em one} loop renormalization of
$(G^a_{\mu\nu})^2$ in QCD in \cite{45}. The upshot is that with ${\cal O}$
inserted in a two point function with a non-zero momentum that Green's function
is in fact a three point function. To apply the {\sc Mincer} algorithm an 
external momentum must be nullified which clearly cannot be that passing 
through the operator. Instead it must be the external momentum associated with 
the field in which the operator is inserted. To determine the three loop result
the only possibilities are the gluon and ghost fields. Inserting in a quark two
point function would require a four loop calculation due to the absence of a 
tree term. Gluon external legs would inevitably lead to an infrared problem 
even at one loop which we are trying to a avoid so we are forced us to consider
a ghost two point function. It transpires, by considering the way the momentum 
is nullified in this case, that spurious infrared infinities cannot arise, in 
much the same way that they do not in the three point function evaluations for 
the earlier coupling constant renormalizations. Hence, we have renormalized 
$\langle c^a(p_1) {\cal O}(p_3) \bar{c}^b(p_2) \rangle$ with {\sc Mincer} using
this procedure where $p_1$ $+$ $p_2$ $+$ $p_3$ $=$ $0$ and $p_1$~$=$~$0$. 
Allowing for the ghost wave function renormalization of the external fields we 
find 
\begin{eqnarray} 
Z_{\cal O} &=& 1 ~+~ \left[ \left( \frac{\alpha}{4} - \frac{35}{12} \right) C_A 
+ \frac{4}{3} T_F \Nf \right] \frac{a}{\epsilon} \nonumber \\
&& +~ \left[ \left( \left( \frac{2765}{288} - \frac{11\alpha}{12} \right) 
C_A^2 ~+~ \frac{16}{9} T_F^2 \Nf^2 ~+~ \left( \frac{\alpha}{3} - \frac{149}{18}
\right) C_A T_F \Nf \right) \frac{1}{\epsilon^2} \right.  \nonumber \\
&& \left. ~~~~~+~ \left( \left( \frac{\alpha^2}{32} + \frac{11\alpha}{32} 
- \frac{449}{96} \right) C_A^2 ~+~ 2 C_F T_F \Nf ~+~ \frac{35}{12} C_A T_F \Nf 
\right) \frac{1}{\epsilon} \right] a^2 \nonumber \\ 
&& +~ \left[ \left( \frac{64}{27} T_F^3 \Nf^3 + \left( \frac{493}{12} 
- \frac{173\alpha}{72} \right) C_A^2 T_F \Nf \right. \right. \nonumber \\
&& \left. \left. ~~~~-~ \left( \frac{154}{9} - \frac{4\alpha}{9} \right) 
C_A T_F^2 \Nf^2 + \left( \frac{3767\alpha}{1152} - \frac{113365}{3456} \right) 
C_A^3 \right) \frac{1}{\epsilon^3} \right. \nonumber \\ 
&& \left. ~~~~+~ \left( \frac{56}{9} C_F T_F^2 \Nf^2 
+ \frac{85}{9} C_A T_F^2 \Nf^2 + \left( \frac{\alpha}{2} - \frac{263}{18} 
\right) C_A C_F T_f \Nf \right. \right. \nonumber \\
&& \left. \left. ~~~~~~~~~~~~+~ \left( \frac{\alpha^2}{24} 
+ \frac{71\alpha}{48} - \frac{5407}{144} \right) C_A^2 T_F \Nf \right. \right.
\nonumber \\ 
&& \left. \left. ~~~~~~~~~~~~+~ \left( \frac{41579}{1152} - \frac{99\alpha}{32}
- \frac{59\alpha^2}{384} - \frac{\alpha^3}{384} \right) C_A^3 \right) 
\frac{1}{\epsilon^2} \right. \nonumber \\  
&& \left. ~~~~+~ \left( -~ \frac{44}{27} C_F T_F^2 \Nf^2 
- \frac{193}{81} C_A T_F^2 \Nf^2 + \left( \frac{415}{108} + 4\zeta(3) \right) 
C_A C_F T_f \Nf \right. \right. \nonumber \\
&& \left. \left. ~~~~~~~~~~~~+~ \left( \frac{5563}{324} 
- \frac{\alpha}{3} - 3\zeta(3) \right) C_A^2 T_F \Nf 
- \frac{2}{3} C_F^2 T_F \Nf \right. \right. \nonumber \\ 
&& \left. \left. ~~~~~~~~~~~~-~ \left( \frac{75607}{5184} 
- \frac{167\alpha}{192} - \frac{101\alpha^2}{768} - \frac{\alpha^3}{128}
- \left( \frac{3}{32} + \frac{\alpha}{8} \right) \zeta(3) \right) C_A^3 \right)
\frac{1}{\epsilon} \right] a^3 \,+\, O(a^4) \nonumber \\  
\label{Zo} 
\end{eqnarray} 
where
\begin{equation} 
{\cal O}_{\mbox{\footnotesize{o}}} ~=~ Z_{\cal O} {\cal O} ~.  
\end{equation} 
Clearly the one and two loop terms agree with the known results for the mass
renormalization in the Curci-Ferrari model itself, \cite{2,5,21,28,29}. 
Moreover, the three loop answer is derived using the {\em same} converter 
routines used for the coupling constant renormalization. Also it is
straightforward to check that like (\ref{Za}) to (\ref{Zwfs}) the triple and 
double pole terms with respect to $\epsilon$ at three loops can be predicted 
from the one and two loop terms and these values in (\ref{Zo}) are in 
{\em exact} agreement for all $\alpha$. Hence we find
\begin{eqnarray} 
\gamma_{\cal O}(a) &=& \left[ 16 T_F \Nf 
+ ( 3\alpha - 35 ) C_A \right] \frac{a}{24} \nonumber \\
&& +~ \left[ 280 C_A T_F \Nf + ( 3\alpha^2 + 33\alpha - 449) C_A^2 
+ 192 C_F T_F \Nf \right] \frac{a^2}{96} \nonumber \\
&& +~ \left[ \left( (2592\alpha + 1944)\zeta(3) + 162\alpha^3 + 2727\alpha^2 
+ 18036\alpha - 302428 \right) C_A^3 \right. \nonumber \\ 
&& \left. ~~~~~-~ \left( 62208\zeta(3) + 6912\alpha - 356032 \right) C_A^2 T_F 
\Nf  + \left( 82944\zeta(3) + 79680 \right) C_A C_F T_F \Nf \right. 
\nonumber \\ 
&& \left. ~~~~~-~ 49408 C_A T_F^2 \Nf^2 - 13824 C_F^2 T_F \Nf 
- 33792 C_F T_F^2 \Nf^2 \right] \frac{a^3}{13824} ~+~ O(a^4) ~. 
\label{gammao}
\end{eqnarray} 
The only other check on this result is the Yang-Mills expression computed in
\cite{2} for $SU(N_c)$ and $\alpha$ $=$ $0$ with the tensor correction method. 
Unfortunately we do not find agreement with \cite{2} for two reasons. First, 
the term of (\ref{gammao}) involving 302428 is 377452 from equation (24) of 
\cite{2} when compared over the same denominator. (The term $1944\zeta(3)$ is 
in agreement with that given in \cite{2}.) However, it is implicit in \cite{2} 
and explicit in \cite{16} that to deduce the corresponding $Z_{\cal O}$ an 
incorrect value of the three loop Landau gauge gluon anomalous dimension has 
been used. Allowing for the correct value, (\ref{Za}), the expressions still 
differ by the quantity $19/24$. We are confident, however, that our result 
(\ref{gammao}) is in fact correct for the following simple reason, which to our
knowledge has not been noted before. If one considers for the moment the one 
and two loop Yang-Mills results for $\alpha$ $=$ $0$ it transpires that the 
coefficients are given by the sum of the gluon and ghost anomalous dimensions 
in the Landau gauge. Indeed it is also apparent that this holds for $\Nf$ 
$\neq$ $0$. Turning to the three loop expression, (\ref{gammao}), it is 
straightforward to observe that the same feature emerges there too. Of course 
such a property could be accidental but it justifies our confidence in the 
veracity of (\ref{gammao}). Moreover, it would also suggest the existence of 
some underlying Slavnov-Taylor identity. If so then $Z_{\cal O}$ is not an 
independent renormalization constant and this would reduce the number of such 
constants required to render (\ref{cflag}) finite. The search for and 
construction of such a Slavnov-Taylor identity is beyond the scope of this 
article but clearly its prior existence would have rendered a certain amount of
our computation unnecessary. However, if constructed it might play a crucial 
role in simplifying the calculation of similar anomalous dimensions in other 
gauges such as the maximal abelian gauge.

Having  deduced  (\ref{gammao}) in the Curci-Ferrari model we can determine
the Landau gauge value which will correspond to the QCD value in the same
gauge. We found  
\begin{eqnarray} 
\left. \gamma_{\cal O}(a) \right|_{\alpha \, = \, 0} &=& \left[ 16 T_F \Nf 
- 35 C_A \right] \frac{a}{24} \nonumber \\
&& +~ \left[ 280 C_A T_F \Nf - 449 C_A^2 + 192 C_F T_F \Nf \right] 
\frac{a^2}{96} \nonumber \\
&& +~ \left[ ( 486\zeta(3) - 75607 ) C_A^3 - \left( 15552\zeta(3) - 89008 
\right) C_A^2 T_F \Nf \right. \nonumber \\ 
&& \left. ~~~~~+ \left( 20736\zeta(3) + 19920 \right) C_A C_F T_F \Nf 
- 12352 C_A T_F^2 \Nf^2 \right. \nonumber \\ 
&& \left. ~~~~~-~ 3456 C_F^2 T_F \Nf - 8448 C_F T_F^2 \Nf^2 \right] 
\frac{a^3}{3456} ~+~ O(a^4) 
\end{eqnarray} 
which also corresponds to the mass renormalization in the Curci-Ferrari model, 
(\ref{cflag}), at three loops and together with (\ref{cfrge}) completes the 
full three loop renormalization. One consequence of our calculation is that we 
can quote the value for the operator ${\cal O}$ in QED. Setting 
$T_F$~$=$~$C_F$~$=$~$1$ and $C_A$ $=$ $0$ we find, from (\ref{gammao}),  
\begin{equation} 
\left. \gamma_{\cal O}(a) \right|_{\mbox{\footnotesize{QED}}} ~=~ \frac{2}{3} 
\Nf a ~+~ 2 \Nf a^2 ~-~ \Nf \left[ 22\Nf ~+~ 9 \right] \frac{a^3}{9} ~+~ O(a^4)
\end{equation}  
for all $\alpha$. 

To conclude with we note that the {\em new} expression for (\ref{gammao}) will
alter the two loop predictions made in \cite{2} for the numerical estimate of
the operator vacuum expectation value generated by the effective potential 
method. Whilst not detracting from the achievement of that tour-de-force it 
would be interesting to repeat those calculations to explore the effect the 
inclusion of quarks has on the vacuum expectation value estimates for the more
realistic case of QCD. This will require (\ref{gammao}) but also needs the full
two loop effective potential of the composite operator with quarks computed in 
the Landau gauge.  

\noindent
{\bf Acknowledgement.} The author acknowledges useful discussions with R.E.
Browne, D. Dudal, K. Knecht and H. Verschelde.  


\end{document}